\documentclass[preprint]{emulateapj} 
\usepackage{natbib}
\usepackage{hyperref}

\shorttitle{Torqued Disks} 
\shortauthors{Spalding \& Batygin} 

\begin{document}

\title{Early Excitation of Spin-Orbit Misalignments in Close-in Planetary Systems}  

\author{Christopher Spalding$^{1}$ \& Konstantin Batygin$^{1,2}$} 
\affil{$^1$Division of Geological and Planetary Sciences, California Institute of Technology, 1200 E. California Blvd., Pasadena, CA 91125}
\affil{$^2$Institute for Theory and Computation, Harvard-Smithsonian Center for Astrophysics, 60 Garden St., Cambridge, MA 02138} 

\email{cspaldin@caltech.edu}

\begin{abstract}
Continued observational characterization of transiting planets that reside in close proximity to their host stars has shown that a substantial fraction of such objects posses orbits that are inclined with respect to the spin axes of their stars. Mounting evidence for the wide-spread nature of this phenomenon has challenged the conventional notion that large-scale orbital transport occurs during the early epochs of planet formation and is accomplished via planet-disk interactions. However, recent work has shown that the excitation of spin-orbit misalignment between protoplanetary nebulae and their host stars can naturally arise from gravitational perturbations in multi-stellar systems as well as magnetic disk-star coupling. In this work, we examine these processes in tandem. We begin with a thorough exploration of the gravitationally-facilitated acquisition of spin-orbit misalignment and analytically show that the entire possible range of misalignments can be trivially reproduced. Moreover, we demonstrate that the observable spin-orbit misalignment only depends on the primordial disk-binary orbit inclination. Subsequently, we augment our treatment by accounting for magnetic torques and show that more exotic dynamical evolution is possible, provided favorable conditions for magnetic tilting. Cumulatively, our results suggest that observed spin-orbit misalignments are fully consistent with disk-driven migration as a dominant mechanism for the origin of close-in planets.  
\end{abstract}

\section{Introduction}

Nearly two decades after the celebrated radial velocity detection of a planet around 51 Peg \citep{mayor,marcy}, the orbital histories of hot Jupiters, (giant planets that reside within $\sim 0.1$ AU of their host stars) remain poorly understood. Conventional planet formation theory \citep{pollack1996} suggests that in-situ formation of hot Jupiters is unlikely, implying that these objects formed beyond the ice-lines of their natal disks (at orbital radii of order $\sim$ a few AU) and subsequently migrated to their present locations. The nature of the dominant migration mechanism, however, remains somewhat elusive.

Broadly speaking, the proposed theoretical mechanisms responsible for delivery of hot Jupiters to close-in radii fall into two categories. The smooth migration category essentially argues that large-scale transport of giant planets is associated with viscous evolution of the disk \citep{lin1996, 2007Icar..191..158M}. More specifically the envisioned scenario suggests that newly-formed giant planets clear out substantial gaps in their protoplanetary disks \citep{goldreich,2011ARA&A..49..195A} and having placed themselves at the gap center (where torques from the inner and outer parts of the disk instantaneously cancel), drift inwards along with the gas. 

A dramatically different story is foretold by the class of violent migration mechanisms. Within the context of this group of descriptions, giant planets initially residing at large orbital radii first attain near-unity eccentricities and eventually get tidally captured onto tighter orbits. The necessary orbital excitations are expected to stem from dynamical processes such as planet-planet scattering \citep{fordrasio,nagasawa,beaugenes}, Kozai resonance with a perturbing binary star \citep{wumurray,fabtremaine,naoz} and secular chaotic excursions \citep{lithwickwu2012}.

From a purely orbital stand point, there appears to be observational evidence for both sets of processes. That is, the existence of a substantial number of (near-) resonant giant exoplanets \citep{2011PASP..123..412W} and direct observations of gaps in protoplanetary disks \citep{2011ApJ...732...42A, 2012ApJ...758L..19H} imply that smooth disk-driven migration is an active process. Simultaneously, the existence of highly eccentric planets such as HD80606b \citep{2009Natur.459..781L} hint at violent migration as a viable option (see however \citealt{2012arXiv1211.0554D}).

In the recent years, observations of the Rossiter-McLaughlin effect \citep{rossiter,mclaugh}, which inform the projected angle between the stellar spin axis and the planetary orbit \citep{2009ApJ...696.1230F}, have placed additional constraints on the hot Jupiter delivery process. Particularly, the data shows that spin-orbit misalignments are generally common within the hot Jupiter population, and the individual angles effectively occupy the entire possible range. Interpreted as relics of hot Jupiter dynamical histories (see however \citealt{2012ApJ...758L...6R} for an alternative view), these observations seemed to strongly favor the category of violent migration mechanisms over disk-driven migration, as spin-orbit misalignments are a natural outcome of the former.

However, a more thorough theoretical analysis shows that spin-orbit alignment is not a necessary feature of disk-driven migration, because a primordial correspondence between the stellar spin axis and the disk angular momentum vector is not in any way guaranteed. To this end, \citet{bate2010} hypothesized that stochastic external forces that act on newly formed protoplanetary disks may give rise to spin-orbit misalignment, while \citet{lai2011} showed that a mismatch between the stellar magnetic axis and the disk orbital angular momentum vector can be further amplified by magnetic torques. 

In a separate effort, \citet{batygin2012} showed that owing to enhanced stellar multiplicity in star-formation environments \citep{ghez1993,kraus2011,marks2012}, secular gravitational perturbations arising from binary companions may torque protoplanetary disks out of alignment with their host stars. This study was subsequently extended by \citet{batad}, who also considered the dissipative effects of accretion, magnetic modulation of stellar rotation as well as the physical evolution of the star and the disk on the excitation of spin-orbit misalignment. Importantly, the latter study demonstrated that the acquisition of stellar obliquity occurs impulsively, via a passage through a secular spin-orbit resonance.

A distinctive prediction made by the disk-torquing model is the existence of coplanar planetary systems, whose orbital angular momentum vectors differ from the spin axes of the host stars. This prediction was recently confirmed observationally by \citet{2013Sci...342..331H} in the Kepler-56 system. Moreover, the statistical analysis of \citet{homies} has shown that the expected spin-orbit misalignment distribution of the disk-torquing model is fully consistent with the observed one. 

Given the aforementioned successes of the the disk-torquing mechanism in resolving the discrepancy between disk-driven migration and spin-orbit misalignments, a thorough examination of the physical process behind the excitation of inclination is warranted. This is the primary aim of the study at hand. Specifically, in this work, we analyze the passage of the star-disk system through a secular spin-orbit resonance, under steady external gravitational perturbations and magnetically-facilitated tiling of the star. The paper is organized as follows. In section 2, we describe the construction of a perturbative model that approximately captures the relevant physics. In section 3, we describe the characteristic behavior exhibited by the model. We conclude and discuss our results in section 4.

\section{Model}

In order to complete the specification of the problem, we must delineate the various ingredients of the model we aim to construct. In particular, these include formulations of the physical evolution of the disk and the central star (section 2.1), magnetically-facilitated tilting of the stellar-spin axis (section 2.2), gravitational interactions between the disk and the binary companion, as well as the gravitational interactions between the central star and the disk (section 2.3). In this work, we opt to neglect the dissipative effects of accretion, as they have been studied within the context of the same problem elsewhere (i.e. \citealt{batad}) and have been found to be unimportant.

We describe our parameterization of the relevant processes below. In interest of minimizing confusion, we adopt the following convention for identically named variables: quantities referring to the disk are primed, those referring to the central star are marked with a tilde, and those referring to the companion star are labeled by an over-bar. Throughout the paper, an emphasis is placed on simplicity inherent to (semi-)analytical approximations, as opposed to precise yet perplexing numerical calculations.

\subsection{Physical Evolution of the Protoplanetary Disk and the Stellar Interior}

Typically quoted lifetimes of protoplanetary disks fall in the range $\sim 1-10$ Myr and almost certainly depend on various parameters such as the host stellar mass \citep{2011ARA&A..49...67W}. We adopt several approximations for the physical evolution of the star and disk, which are specific to Sun-like stars, which host the best observationally characterized hot Jupiters. While generally difficult to accurately parameterize, the disk mass can be taken to evolve as \citep{laughlin2004}: 
\begin{equation}
M_{\rm{disk}} = \frac{M_{\rm{disk}}^0}{1 + t/\tau_{\rm{disk}}}. 
\end{equation}
Interpreting the time derivative of $M_{\rm{disk}}$ to represent the accretionary flow, following \citet{batad} we find that the initial disk mass, $M_{\rm{disk}}^0 = 5 \times 10^{-2} M_{\odot}$ and evaporation timescale $\tau_{\rm{disk}} = 5 \times 10^{-1}$ Myr provide an acceptable match to the observations \citep{hart2008, herczeg2008, lynne2008}. 

For simplicity, we model the interior structure of the central star with a polytrope of index $\xi=3/2$ (appropriate for a fully convective object; \citealt{chandra}). A polytropic body of this index is characterized by a specific moment of inertia $I = 0.21$ and a Love number (twice the apsidal motion constant) of $k_2 = 0.14$. Because T-Tauri stars derive a dominant fraction of their luminosity from gravitational contraction, we adopt the following expression for the radiative loss of binding energy \citep{hansenkawal}:
\begin{equation}
- 4 \pi R_\star^2 \sigma T_{\rm{eff}}^4 =  \left( \frac{3}{5-\xi} \right) {G M_\star^2 \over 2 R_\star^2} {d R_\star \over dt} . 
\label{drdt}
\end{equation}

Equation (\ref{drdt}) effectively dictates the process of Kelvin-Helmhotz contraction, and is satisfied by the solution:
\begin{equation}
R_{\star} = (R_{\star}^0) \left[1+ \left( \frac{5-\xi}{3} \right) \frac{24 \pi \sigma T_{\rm{eff}}^4}{G M_{\star} (R_{\star}^0)^3} t \right]^{-1/3}.
\label{roft}
\end{equation}
A good match to the numerical evolutionary track of \citet{siess} for a $M_{\star} = 1 M_{\odot}$ star can be obtained by assuming an initial radius of $R_{\star}^0 \simeq 4 R_{\odot}$ and an effective temperature of $T_{\rm{eff}} = 4100$K.

\subsection{Magnetic Torques}

In order to model the magnetic disk-star interactions, we consider a T Tauri star possessing a pure dipole magnetic field, whose north pole is aligned with the stellar spin axis. In the region of interest (i.e. in the domain of the disk), the field is current-free and can be expressed as a gradient of a scalar potential:
\begin{equation}
\vec{B}_{\rm{dip}}=-\vec{\nabla} V.
\end{equation}
To retain generality, we take the field to be tilted at an angle $\beta_i$ with respect to the disk plane into a direction specified by an azimuthal angle $\tilde{\phi}_i$:
\begin{eqnarray}
V&=&B_{\star}R_\star \bigg(\frac{R_\star}{r}\bigg)^2 \bigg[ P_0^1(\cos(\tilde{\theta})) \cos(\beta_i)\nonumber \\
&-&\sin(\beta_i)\bigg(\sin(\tilde{\phi}_i)\sin(\tilde{\phi}) \nonumber \\
&+&  \cos(\tilde{\phi}_i)\text{cos}(\tilde{\phi})\bigg)P_1^1(\text{cos}(\tilde{\theta})) \bigg]
\end{eqnarray}
where $B_{\star}$ is the stellar surface field and $P_l^m$ are associated Legendre polynomials.

If we assume the disk to be circular and Keplerian, there exists a radius, $a'_{\rm{co}}=(G\,M_{\star}/\omega^2)^{1/3}$ at which the mean motion of the disk material, $n'$, equals the spin rate of the stellar magnetic field, $\omega$. At larger and smaller radii, Keplerian shear will give rise to relative fluid velocity with respect to the stellar rotation. Accordingly, as a result of thermal ionization of alkali metals in the disk \citep{1983ApJ...264..485D}, the magnetic field will be dragged azimuthally by differential rotation, whilst slipping backwards diffusively \citep{1992MNRAS.259P..23L}. Following \citet{1996MNRAS.280..458A} we parameterize the magnitude of the azimuthally-induced field $B_\phi$ as a fraction, $\gamma=B_{\phi}/B_z$ of the vertical component of the dipole field $B_z$.

As shown by \citet{2000MNRAS.317..273A} and \citet{2002ApJ...565.1191U}, beyond a critical value of $\gamma \simeq 1$, field lines are stretched to a sufficient degree to reconnect and transition from a closed to an open topology. Thus, the condition $|\gamma | \lesssim 1$  defines a magnetically-connected region within the disk with $\hat{a}'_{\rm{in}}<a'<\hat{a}'_{\rm{out}}$. Outside of this region, we assume there to be no appreciable magnetic coupling to the disk. 

The radial profile of $\gamma$ is determined by the magnetic diffusivity of the disk, which in turn may be represented by the dimensionless parameter \citep{mattpud}:
\begin{equation}
\zeta\,=\,\frac{\alpha}{P_m}\frac{h}{a'},
\end{equation}
where $\alpha$ is the disk viscosity parameter introduced  by \citet{1973A&A....24..337S}, $P_m$ is the Magnetic Prandtl number, and $h$ is the scale height of the disk. As argued by \citet{mattpud,2005MNRAS.356..167M}, any realistic choice of parameters yields $\zeta\,\leq\,10^{-2}$, which is the value we adopt throughout our work here. In terms of $\zeta$, a steady state magnetic twist angle may be expressed as \citep{2002ApJ...565.1191U}
\begin{equation}
\gamma\,=\,\frac{(a'/a'_{\rm{co}})^{3/2}-1}{\zeta}.
\label{gammayo}
\end{equation}
For our adopted value of $\zeta$ this, so-called, magnetically-connected region does not diverge from the corotation radius by more than $\sim\,1\%$. 

The above discussion highlights a crucial aspect of the magnetic star-disk interaction, which is discussed in detail by \citet{mattpud,2005MNRAS.356..167M}. If the disk is truncated at $a'_{\rm{in}}\,>\,\hat{a}'_{\rm{out}}$, then there is no magnetically-connected region within the disk. The picture is slightly more complicated for the case where $a'_{\rm{in}}\,<\hat{a}'_{\rm{in}}$, as one may speculate that magnetic effects arising from differential rotation outside $a'_{\rm{co}}$ may cancel those associated with differential rotation inside $a'_{\rm{co}}$ to first order. In all of the following work, we circumvent these issues by assuming a disk-locked condition \citep{shu1994,subu} where $a'_{\rm{in}}\,=\,a'_{\rm{co}}$, but add a cautionary note that this assumption may lead to somewhat overly favorable results. 

In order to derive the analytical form of the torques, we take a similar approach to that of \citet{lai2011}, and assume the disk to be razor thin. The disk current loops are envisioned to follow the magnetic field lines in a force-free fashion (see e.g.
\citealt{2009ApJ...703.1863K, 2013A&A...550A..99Z}) and connect onto the stellar surface. Accordingly, the induced azimuthal magnetic field arises from a radial current within the disk \citep{1999ApJ...524.1030L}. 

The magnitude of the radial current is calculated using Amp$\grave{\rm{e}}$re's Law \citep{1998clel.book.....J} in the form:
\begin{equation}
 \int_C \vec{B}\cdot d\vec{l} = \mu_0  \int \int_A \vec{J}\cdot d\vec{S}  
\end{equation}
where $d\vec{l}$ is a vector path length, $d\vec{S}$ is a vector area element and $C$ is a loop encompassing the surface $A$. $\vec{J}$ is the (induced) current density within the disk. Because the induced field is entirely toroidal, we can integrate the left hand side along an azimuthal loop around the disk. This yields:
\begin{equation}
4\pi a' B_\phi = 2 \pi \mu_0 a' K_{r},
\end{equation}
where $K_{r} = \int J_{r} dz = 2B_\phi/\mu_0$ is the inward radial surface current. 

Combining this expression with equation (\ref{gammayo}), we obtain:
\begin{equation}
K_r\,=\frac{2 B_z}{\mu_0} \left[ \frac{(a'/a'_{\rm{co}})^{3/2}-1}{\zeta}\right].
\end{equation}
With an expression for the induced current at hand, we immediately arrive at an expression for the associated Lorentz torque, considering the induced current to interact only with the stellar dipole field:
\begin{equation}
\vec{\tau}_L=(a'\,\vec{\hat{\rho}}\,)\times (K_r\,\vec{\hat{\rho}} \times \vec{B}_{\rm{dip}}).
\end{equation}
In the above expression, $\hat{\rho}$ is the radial unit vector in the plane of the disk.  

At this point, in order to cast the magnetic torques into a usable form, we project $\vec{\tau}_L$ onto each of the Cartesian axes in the disk's frame 
and subsequently integrate over the entire magnetically-connected region: \begin{equation}
\tau_{i'}=\int_0^{2\pi} \int_{a'_{\rm{co}}}^{\hat{a}'_{\rm{out}}}\vec{\tau}_L \cdot \vec{\hat{x}}_{i'}\, \rho\, d\rho\, d\phi
\end{equation}
where the subscript $i'$ represents the Cartesian axes in the disk's frame. With the variables and parameters given above, these torques evaluate to:
\begin{eqnarray}
\tau_{x'}= \bigg( \frac{2 \pi B_{\star}^2\,R_{\star}^6\, \zeta\, \sin(\beta_i)\, \cos(\beta_i)}{3\mu_0 (1+\zeta)^2 (a'_{\rm{co}})^3} \bigg)\,\cos(\tilde{\phi}_i)
\end{eqnarray}
\begin{eqnarray}
\tau_{y'}=\bigg(\frac{2 \pi B_{\star}^2\,R_\star^6\, \zeta\, \sin(\beta_i)\,  \cos(\beta_i)}{3\mu_0 (1+\zeta)^2 (a'_{\rm{co}})^3}\bigg)\,\sin(\tilde{\phi}_i)
\end{eqnarray}
\begin{eqnarray}
\tau_{z'}=\bigg(\frac{4\pi B_{\star}^2R_\star^6\, \zeta\, \cos^2(\beta_i)}{3\mu_0(1+\zeta)^2 (a'_{\rm{co}})^3}\bigg).
\end{eqnarray}

Note that for $\beta_i > \pi /2$, the star and disk spin in opposite directions and so the concept of a corotation radius loses meaning. Accordingly, under the assumptions of our model, no magnetically connected region exists as presented above. Indeed, it is unclear how the magnetic field would interact with the disk in such a case and consequently, additional torques or accretional processes may have been omitted. For the purposes of our \text{idealised} model, we incorporate the loss of a magnetically connected region by artifically forcing the torques to equal zero for angles of $\beta_i> \pi/2$. Mathematically, this is done by multiplying the torques by an approximation to a step function $\mathcal{S}(\beta_i)$ given by:
\begin{equation}
\mathcal{S}(\beta_i)=1-\bigg(\frac{\pi/2+ \arctan \big(\frac{\beta_i-\pi/2}{\ell}\big)}{\pi}\bigg),
\end{equation}
where $\ell = 10^{-4}$. The importance of such a term becomes apparent once the torques are coupled to gravity and inclinations above 90 degrees are naturally attained.

Angular momentum transport among neighboring annuli of the disk is facilitated by propagation of bending waves \citep{2011MNRAS.412.2799F} as well as disk self-gravity \citep{batygin2011} and generally occurs on a much shorter timescale than magnetic tilting of the host star. Taking advantage of this, in our analyses we assume that the effective moment of inertia of the disk around all axes is much greater than that of the star, allowing us to ignore any torques from the star on the disk and to consider $-\tau_{i'}$ as a back-reaction on the star's dynamics.

As will become apparent below, it is beneficial to carry out all calculations in the frame of a distant, binary companion to the central star. As such, we follow the approach of \citet{2014arXiv1401.4131P} and define Euler angles within the binary frame related to the nutation, precession and rotation of the rigid body while assuming exclusively principal axis rotation (this is an excellent approximation for a T-Tauri star, spinning at a period of 1-10$\,$days). Specifically, $\tilde{\beta}$ is the angle between the central star's spin axis and the binary orbit normal; $\tilde{\Omega}$ is the longitude of ascending node of the star in the binary frame where $\tilde{\Omega} = 0$ implies collinear disk and stellar lines of nodes; and the third Euler angle $\varphi$ is the angle through which the star rotates as it spins ($\varphi$ only enters the equations as a rate of change: $\dot{\varphi} = \omega$). 

The equations for the evolution of $\tilde{\beta}$ and $\tilde{\Omega}$, adapted from \citet{2014arXiv1401.4131P} are:
\begin{eqnarray}
\label{pealeequns}
\frac{d\tilde{\beta}}{dt}&=&-\frac{1}{\omega} \bigg[\cos(\tilde{\beta})(-N_{\bar{x}} \sin(\tilde{\Omega})+N_{\bar{y}} \cos(\tilde{\Omega}))\nonumber\\
&+&N_{\bar{z}} \sin(\tilde{\beta}))\bigg],
\end{eqnarray}
\begin{equation}
\frac{d\tilde{\Omega}}{dt}=-\frac{1}{\omega\, \sin(\tilde{\beta})}\bigg[N_{\bar{x}} \cos(\tilde{\Omega})+N_{\bar{y}} \sin(\tilde{\Omega})\bigg],
\end{equation}
where $N_{\bar{i}}$ are projected torques. Note that by fixing the disk's longitude of ascending node at $\Omega' = 0$, we have implicitly placed ourselves into a frame coprecessing with the disk's angular momentum vector. The effect of precession shall be included within the gravitational part of the equations and we need not retain it here. 

Throughout the pre-main-sequence, we assume a constant rotation rate of the host star. Although stellar rotation is almost certainly modulated by the presence of the disk \citep{herb2007,affer,bouvier2013}, the present lack of detailed understanding of the physical processes behind rotational breaking render this assumption reasonable (see \citealt{gallet}).

The projected quantities $N_{\bar{i}}$ are directly related to the torques calculated above, although the components of the torques in the disk frame, $-\tau_{i'}$, must first be projected onto the Cartesian axes in the binary frame. Such a projection constitutes a simple rotation of co-ordinates because, as discussed below, the disk-binary inclination is a constant of motion. The rotation angle is fixed at some prescribed angle, $\beta'$, anti-clockwise about the x-axis. As such, the components, $N_{\bar{i}}$ are given in terms of $\tau_i$ by:
\begin{equation}
N_{\bar{x}}=-\tau_{x'}/(I M_{\star} R_{\star}^2),
\end{equation}
\begin{equation}
N_{\bar{y}}=-(\cos(\beta')\tau_{y'}-\tau_{z'} \sin(\beta'))/(I M_{\star} R_{\star}^2),
\end{equation}
\begin{equation}
N_{\bar{z}}=-(\cos(\beta')\tau_{x'}+\tau_{y'}\sin(\beta'))/(I M_{\star} R_{\star}^2).
\end{equation} 

The above equations can be used to analyze the dynamics of the central star owing to its magnetic field interacting with its protoplanetary disk. It is noteworthy that we have made no mathematical assumption of small angles, but physically, we have not taken into account the changes in the parameterized geometry of the problem that arise when mutual disk-star inclinations approach $\beta_i \rightarrow \pi/2$. Without a doubt, future calculations should consider this effect, as we expect changes in magnetic field geometry to affect the detailed nature of the relevant torques. At the same time, it is plausible that despite quantitative differences, the overall qualitative picture envisioned within the context of more precise calculations will not be in stark contrast to the parameterized model employed here.

\subsection{Gravitational Torques}

\subsubsection{Binary Star - Disk Interactions}

In this section, we calculate the gravitational response of a disk to a distant, binary companion, whose orbit is taken to lie in the reference plane. We derive a Hamiltonian for the disk subject to perturbations from the companion, working under the secular approximation. In other words, we assume that the disk's outer radius $a_{\rm{out}}^\prime$ is sufficiently small compared to the binary orbit's semi-major axis $\bar{a}$ that no meaningful commensurabilities between the orbital motions exist. 

The Gaussian averaging method (see Ch. 7 of \citealt{md2000}) dictates that in the aforementioned regime, the (orbit-averaged) treatment of the gravitational interactions of the disk-companion system is mathematically equivalent to considering the companion to be a circular ring with line density $\lambda=\bar{M}/(2\pi \bar{a})$ and the disk as an infinite series of annular wires at every radius between $a'_{\rm{in}}$ and $a'_{\rm{out}}$ \citep{md2000,morby2012}. 

It is well known that within the secular framework, the semi-major axes are constants of motion \citep{morbybook}. Consequently, the Keplerian contribution to the Hamiltonian can be dropped, rendering the Hamiltonian of this set up, simply the total gravitational potential energy $\mathcal{U}$ possessed by the disk in the field of the companion ring: 
\begin{equation}
\label{UG}
\mathcal{U}=-\int_{\rm{disk}} \int_{\rm{ring}}\frac{G}{r}\,dM_{\rm{disk}}\,dM_{\rm{ring}},
\end{equation}
where $r$ is the separation between two mass elements $dM_{\rm{ring}}$ (companion), and $dM_{\rm{disk}}$ (disk). The integral is carried out over all angles ($\bar{\phi}$) within the ring and over all radii ($a'$) and angles ($\phi'$) in the disk. 
The evaluation of $r(\phi',a',\bar{\phi})$ is a purely geometric problem and can be simplified by approximating $a_{\rm{out}}^\prime/ \bar{a}\ll1$. Under such an approximation, we expand $r$ to second order in equation (\ref{UG}) (first order terms are axisymmetric and therefore cancel out). 

In order to compute the integral (\ref{UG}), we must specify the disk surface density profile, $\Sigma$. For definitiveness, in this work we shall follow \citep{1963MNRAS.126..553M, batygin2012} and adopt
\begin{eqnarray}
\Sigma = \Sigma_0 \bigg( \frac{a'_0}{a'} \bigg),
\label{SD}
\end{eqnarray}
where $\Sigma_0$ is the surface density at semi-major axis $a'_0$. We note that the passive disk model of \citet{1997ApJ...490..368C} is characterized by a very similar power-law index: $\Sigma \propto (a')^{-15/14}$ \citep{2012arXiv1212.2217R}. 

With this prescription, equation (\ref{UG}) becomes:
\begin{equation}
\mathcal{U}=-\int_{a'_{\rm{in}}}^{a'_{\rm{out}}}\int_0^{2\pi}\int_0^{2\pi}\frac{G}{r}\Sigma_0\,a'_0\,\frac{\bar{M}}{2\pi}\,d\bar{\phi}\,d\phi'\,da'.
\end{equation} 
Noting that $a_{\rm{in}}^\prime\ll a_{\rm{out}}^\prime$, we arrive at the expression for the Hamiltonian. 

Switching to canonically conjugated variables, we introduce the scaled Poincar$\acute{\rm{e}}$ action-angle coordinates
\begin{equation}
\label{scalepoinc}
Z'=1-\cos(\beta') \ \ \ \ \ \ \ z' = -\Omega'.
\end{equation} 
This definition of the coordinates differs from the standard definition (see Ch. 2 of \citealt{md2000}) in that at each disk annulus, the standard definition multiplies $Z'$ by $d\Lambda = dm' \sqrt{G M_{\star} a'}$ where $dm'=2 \pi \Sigma_0 a'_0 da'$ is the mass of the annulus. Thus, for the variables (\ref{scalepoinc}) to remain canonical, we must also scale the Hamiltonian itself in a corresponding manner:
\begin{eqnarray}
\breve{\mathcal{U}}&=&\frac{\mathcal{U}}{2 \pi \int_{a'_{\rm{in}}}^{a'_{\rm{out}}} \Sigma_0 a'_0 \sqrt{G M_{\star} a'} da'} \nonumber \\
&=& \frac{3 n'_{\rm{out}}}{8} \frac{\bar{M}}{M_{\star}} \left(\frac{a'_{\rm{out}}}{\bar{a}} \right)^3 \left[ Z'-\frac{Z'^2}{2} \right].
\label{HUpsilon}
\end{eqnarray}
This expression agrees with the fourth-order Lagrange-Laplace expansion of the disturbing function \citep{md2000}, where Laplace coefficients are replaced with their leading order hypergeometric series approximations \citep{batad}.

The crucial result here is that $\breve{\mathcal{U}}$ does not depend on $z'$, meaning that the disk inclination is exactly preserved in the binary frame:
\begin{eqnarray}
\frac{dZ'}{dt}=-\frac{\partial \breve{\mathcal{U}}}{\partial z'}=0
\end{eqnarray}
while the precession rate depends on both the companion semi-major axis and mass. As shown by \citet{batygin2012} via a different approach, the constancy of disk-star inclination holds even if an eccentric companion is considered. In this case, however, the precession rate is enhanced by a factor of $(1 + 3 \bar{e}/2)$.

\subsubsection{Disk - Central Star Interactions}

As already mentioned above, the spin rates of classical T-Tauri stars fall within the characteristic range of $1-10$ days \citep{herb2007,bouvier2013}. This results in substantial rotational deformation of young stars. To an excellent approximation, the dynamical response of a spheroidal star to the gravitational potential of the disk can be modeled by considering the inertially equivalent configuration whereby a ring of mass
\begin{equation}
\tilde{m} = \left[ 
\frac{3 M_{\star}^2 \omega^2 R_{\star}^3 I^4}{4 G k_2 } \right]^{1/3},
\end{equation}
orbits the star with semi-major axis
\begin{equation}
\tilde{a} = \left[
\frac{16 \omega^2 k_2^2 R_{\star}^6}{9 I^2 G M_{\star}} \right]^{1/3}.
\label{astaryo}
\end{equation} 
Within the context of this picture, the standard perturbation techniques of celestial mechanics can be applied \citep{md2000,morbybook}.

In the exploratory study of \citet{batad}, the gravitational torques were computed using a low mutual inclination approximation to the true potential. This simplification is limiting, especially on the quantitative level, as it forces the topology of the dynamical portrait to be that of the second fundamental model for resonance \citep{henrardlamaitre}, for all choices of parameters. In this work, we shall place no restrictions on the mutual inclination and adopt the series of \citet{1962AJ.....67..300K} as a representation of the potential, which utilizes the semi-major axis ratio $(\tilde{a}/a')$ as an expansion parameter. Provided the smallness of the semi-major axis ratio inherent to our problem ($\tilde{a}/a' \lesssim \mathcal{O}(10^{-1})$), an octupole-order expansion suffices our needs. 

Written in terms of scaled canonical Poincar$\acute{\rm{e}}$ action-angle coordinates (\ref{scalepoinc}), the Hamiltonian that governs the dynamics of the stellar spin-axis\footnote{In unanimity with the above treatment, the back-reaction of the star onto the disk is ignored.} under the gravitational influence of an infinitesimal wire of mass $dm'$ reads:
\begin{eqnarray}
d\mathcal{H} &=& \frac{1}{16}\sqrt{\frac{G M_\star}{a'^3}}\frac{dm'}{M_\star} \left(\frac{\tilde{a}}{a'}\right)^{3/2} \bigg[ \big(2 - 6 \tilde{Z} + 3 \tilde{Z}^2 \big) \nonumber \\ \nonumber \\
&\times& \big(2 - 6 Z' + 3 Z'^2\big)+ 12 \left(\sqrt{\tilde{Z}(2-\tilde{Z})} - \sqrt{\tilde{Z}^3(2-\tilde{Z})} \right) \nonumber \\
&\times& \left(\sqrt{\tilde{Z}(2-\tilde{Z})} - \sqrt{\tilde{Z}^3(2-\tilde{Z})} \right) \cos(\tilde{z} - z') \nonumber \\
&+& 3 \tilde{Z} Z' \big(\tilde{Z} - 2 \big) \big(Z' - 2 \big) \cos\big(2(\tilde{z} - z')\big) \bigg].
\label{Hwire}
\end{eqnarray}
As above, to obtain an expression for the Hamiltonian that incorporates the effect of the entire disk, we imagine the disk to be composed of a series of such aforementioned wires and integrate: 
\begin{eqnarray}
\mathcal{H} &=& \int d\mathcal{H}.
\label{Hint}
\end{eqnarray}
Recalling that the mass of each individual wire comprising the disk is $dm'=2 \pi \Sigma_0 a'_0 da'$, we note that the integral (\ref{Hint}) runs with respect to the disk semi-major axis. 

Because of the stiff dependence of equation (\ref{Hwire}) on $a'$, the integral (\ref{Hint}) is not explicitly sensitive to the location of the outer edge of the disk. Rather, it depends upon the disk's total mass, provided that the former is substantial (i.e. 10s of AU; \citealt{anderson}). In turn, the disk's mass is predominantly set by the disk's size, $a'_{\rm{out}}$:
\begin{equation}
M_{\rm{disk}} = 2 \pi \int_{a'_{\rm{in}}}^{a'_{\rm{out}}} a' \Sigma_{0} \left( \frac{a'_0}{a'} \right) da' \simeq 2 \pi \Sigma_{0} a'_0 a'_{\rm{out}}.
\end{equation}
In addition to the disk's physical properties, we must also prescribe its dynamical behavior to complete the specification of the problem. As already mentioned above, because the Hamiltonian (\ref{HUpsilon}) is independent of the angles (i.e. is a Birkhoff normal form), the disk inclination with respect to the binary orbital plane (and therefore $Z'$) is conserved, while the disk's nodal precession rate, $\nu = dz'/dt$, is given by 
\begin{equation}
\nu = \frac{\partial \breve{\mathcal{U}}}{\partial Z'} = \frac{3 n'_{\rm{out}}}{8} \frac{\bar{M}}{M_{\star}} \left(\frac{a'_{\rm{out}}}{\bar{a}} \right)^3 \bigg[ \,1- Z' \bigg] .
\end{equation}
Consequently, $\mathcal{H}$ represents a non-autonomous one degree of freedom system. 

Because the time-dependence inherent to the problem at hand is particularly simple ($z' = \nu t$), $\mathcal{H}$ can be made autonomous by employing a canonical transformation arising from the following generating function of the second kind \citep{1950clme.book.....G}:
\begin{equation}
\mathbb{G}_2 = (\tilde{z}-\nu t) \Phi,
\end{equation}
where $\phi = (\tilde{z}-\nu t)$ is the new angle and the new momentum is related to the old one through:
\begin{equation}
\tilde{Z} = \frac{\partial \mathbb{G}_2}{\partial \tilde{z}} = \Phi.
\end{equation}
Accordingly, the Hamiltonian itself is transformed as follows \citep{1983rsm..book.....L}:
\begin{equation}
\mathcal{K} = \mathcal{H} - \frac{\partial \mathbb{G}_2}{\partial t}.
\end{equation}
Having removed explicit time dependence, we additionally scale the Hamiltonian by $\nu$, which introduces a single dimensionless number; the ``resonance proximity parameter," that encompasses the physical properties of the system. An explicit expression for $\tilde{\delta}$ reads:
\begin{eqnarray}\label{delta}
\tilde{\delta}=\frac{3}{8}\left(\frac{{n'}_{\rm{in}}^2}{\omega \nu}  \frac{M_{\rm{disk}}}{M_{\star}} \frac{a'_{\rm{in}}}{a'_{\rm{out}}} \right).
\end{eqnarray}

Following the transformations described above, the Hamiltonian takes on the following form:
\begin{eqnarray}
\mathcal{K} &=& - \Phi + \frac{\tilde{\delta}}{12} \bigg[ 3 \big(\Phi -2 \big) \Phi - 3 \big(2+3(\Phi -2)\big) \Phi \cos^2(\beta') \nonumber \\ 
&+& 6 \sin(2 \beta') \big(\Phi -1\big) \sqrt{(2-\Phi) \Phi} \cos(\phi) + 3 \sin^2(\beta')  \nonumber \\ 
&\times& \big(\Phi -2\big) \Phi \cos(2 \phi) \bigg].
\label{HammyK}
\end{eqnarray}

\begin{figure}
\includegraphics[width=1\columnwidth]{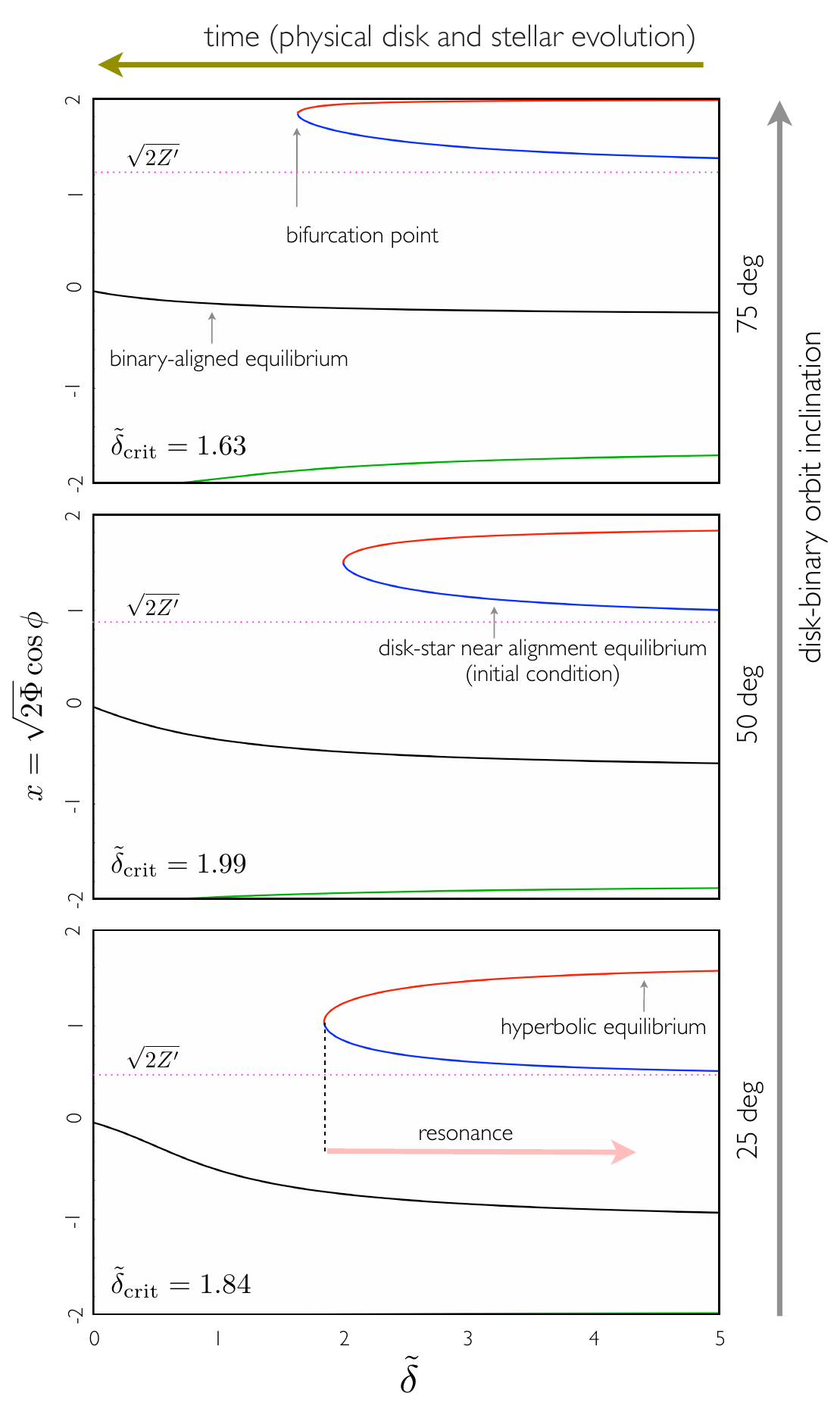}
\caption{Equilibria of the Hamiltonian (\ref{HammyK}) as a function of the resonance proximity parameter $\tilde{\delta}$ (see equation \ref{delta}). The three panels correspond to different choices of disk-binary inclination, namely $\beta' = 25 \deg, \beta' = 50 \deg,$ and $\beta' = 75 \deg$. The equilibria depicted in black, blue, and green lines are stable, while that shown as a red line is unstable. As $\tilde{\delta}\rightarrow\tilde{\delta}_{\rm{crit}}$, two of the four solution merge onto a single unstable equilibrium. On the contrary, as $\tilde{\delta}\rightarrow\infty$, a stable equilibrium point approaches perfect alignment with the disk (shown as a dashed line).} 
\label{equilibplot}
\end{figure}

\section{Results}

With a theoretical model in place, let us begin our exploration of the acquisition of spin-orbit misalignments in an idealized limit. That is, we begin by neglecting magnetic torques and assuming adiabaticity. 

\subsection{Purely Gravitational Excitation}

Although the Hamiltonian (\ref{HammyK}) does not exhibit explicit time-dependence, it does possess implicit time-dependence through the evolution of resonance proximity parameter, $\tilde{\delta}$. As discussed in \citet{batad}, the time-dependent variation in $\tilde{\delta}$ is primarily brought about as a result of disk mass loss and the physical evolution of the host star (via $\tilde{n}$). While both of these processes can be quite complex, for our purposes, it suffices to note that for any reasonable choice of parameters, $\tilde{\delta} \rightarrow 0$ as the system approaches main sequence. 

What are the consequences of the changes in the value of $\tilde{\delta}$? Preliminary progress towards understanding this question can be made by studying the equilibria of the Hamiltonian (\ref{HammyK}). To do so, it is particularly convenient to switch to cartesian coordinates defined as 
\begin{eqnarray}
x= \sqrt{2 \Phi} \cos(\phi) \ \ \ \ \ y= \sqrt{2 \Phi} \sin(\phi).
\end{eqnarray}
The angular dependence of $\mathcal{K}$ shows that all equilibria will have $\phi = 0$, also implying $y = 0$ \citep{md2000}. The equilibrium values of $x$ are shown as functions of $\tilde{\delta}$ in Figure (\ref{equilibplot}) for three choices of disk-binary orbit inclination.

Depending on the value of $\tilde{\delta}$, the Hamiltonian $\mathcal{K}$ possesses between two and four fixed points. Generally, for $\tilde{\delta}<1$ two stable (elliptic) fixed points exist (shown in black and green), while four fixed points (one of which is stable (shown in blue) and the other is unstable i.e. hyperbolic (shown in red)) are guaranteed for $\tilde{\delta}>2$. There exists a critical bifurcation value $1 \leqslant \tilde{\delta}_{\rm{crit}} \leqslant 2$ where the number of fixed points is three and the stable and unstable fixed points merge into a single unstable equilibrium \citep{henrardlamaitre,peale1986}. As shown in Figure (\ref{equilibplot}), $\tilde{\delta}_{\rm{crit}}$ depends on the disk-binary orbit inclination: it changes smoothly from $\tilde{\delta}_{\rm{crit}} = 1$ at $\beta' = 0 \deg$ to $\tilde{\delta}_{\rm{crit}} = 2$ at $\beta' = 45 \deg$, and back to $\tilde{\delta}_{\rm{crit}} = 1$ at $\beta' = 90 \deg$.

\begin{figure*}
\includegraphics[width=1\textwidth]{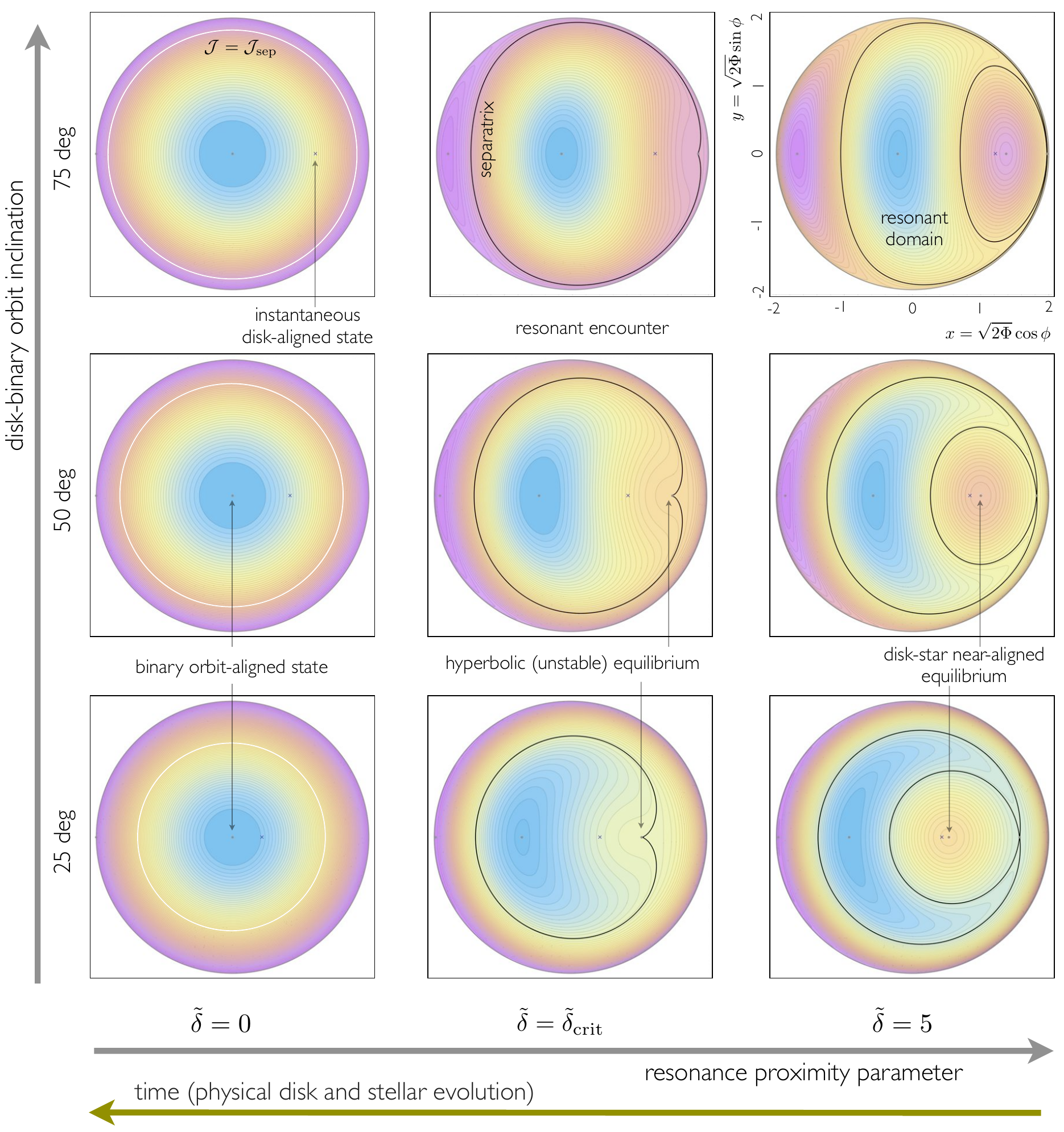}
\caption{Phase-space portraits of the Hamiltonian (\ref{HammyK}) at different values of the resonance proximity parameter $\tilde{\delta}$ and disk-binary inclination. The colors represent the value of the Hamiltonian at each contour. In all panels, the equilibria of the Hamiltonian are shown as gray dots. The instantaneous disk aligned state is depicted with a small $\times$ symbol. Note that for $\tilde{\delta}$ well above $\tilde{\delta}_{\rm{crit}}$, there exist an equilibrium point in close proximity (but not exactly corresponding to) the disk aligned state. The separatrix is shown as a black curve for $\tilde{\delta} \geqslant \tilde{\delta}_{\rm{crit}}$. On panels corresponding to $\tilde{\delta} = 0$, a white circular orbit that occupies the same phase-space area as the separatrix at $\tilde{\delta} = \tilde{\delta}_{\rm{crit}}$ is shown.} 
\label{pspace}
\end{figure*}

Physically, $\tilde{\delta}$ represents the ratio of the characteristic precession rates of the star's and disk's angular momentum vectors. It is generally safe to assume that this ratio is well above unity at the epoch when the system emerges from the embedded stage (see e.g. \citealt{2011ARA&A..49...67W}). Moreover, it can be argued that dissipative processes such as accretion will force the stellar spin axis to evolve towards the equilibrium point closest to alignment with the disk's angular momentum vector (see \citealt{2011CeMDA.111..219B} for a related discussion). Provided that the changes in $\tilde{\delta}$ are slow compared to the characteristic precession period of the star, adiabatic theory dictates that the (null) phase-space area (i.e. the adiabatic invariant $\mathcal{J}$) associated with the orbit of the stellar spin axis must be approximately conserved \citep{henrardbible}: $\mathcal{J}_{\rm{eq}} = 0$. Consequently, as $\tilde{\delta}$ decreases in time the stellar spin axis will reside on the equilibrium solution shown in blue on Figure (\ref{equilibplot}). However, once the evolutionary track of the system reaches $\tilde{\delta} = \tilde{\delta}_{\rm{crit}}$, the associated equilibrium becomes unstable. 

To understand the dynamical evolution of the stellar spin axis beyond the aforementioned adiabatic trailing phase, it is useful to consider the geometry of the Hamiltonian. For the three choices of inclination depicted in Figure (\ref{equilibplot}), Figure (\ref{pspace}) shows a series of phase-space portraits of $\mathcal{K}$ in cartesian coordinates. Specifically, the panels of the Figure (\ref{pspace}) depict snap-shots of the Hamiltonian flow at $\tilde{\delta} = 5$, $\tilde{\delta} = \tilde{\delta}_{\rm{crit}}$, and $\tilde{\delta} = 0$. Here, the equilibrium points of $\mathcal{K}$ are shown as gray dots, while the separatrix (i.e. homoclinic orbit) associated with the secular spin-orbit resonance is depicted as a black curve where it exists (i.e. $\tilde{\delta} \geqslant \tilde{\delta}_{\rm{crit}}$).

Qualitatively, the following picture holds. As long as $\tilde{\delta} > \tilde{\delta}_{\rm{crit}}$, the system remains adiabatically frozen on the equilibrium point contained in the inner circulation zone of the separatrix. As $\tilde{\delta} \rightarrow \tilde{\delta}_{\rm{crit}}$, the phase space area associated with the inner circulation zone shrinks and eventually the equilibrium point on which the stellar spin-axis resides is invaded by the separatrix. Because the separatrix is characterized by an infinite period, the adiabatic principle inevitably breaks down and the conservation of $\mathcal{J}$ is momentarily violated. However, immediately after the encounter, the separatrix turns into a regular circulatory orbit and the system returns into the realm of adiabatic theory \citep{1984CeMec..32..127B, 1991pscn.proc..193H, 2013A&A...556A..28B}. As such, for all subsequent evolution, the value of the adiabatic invariant remains equivalent to that of the separatrix, evaluated at the critical resonance proximity parameter \citep{peale1986}. 

\begin{figure}
\includegraphics[width=0.95\columnwidth]{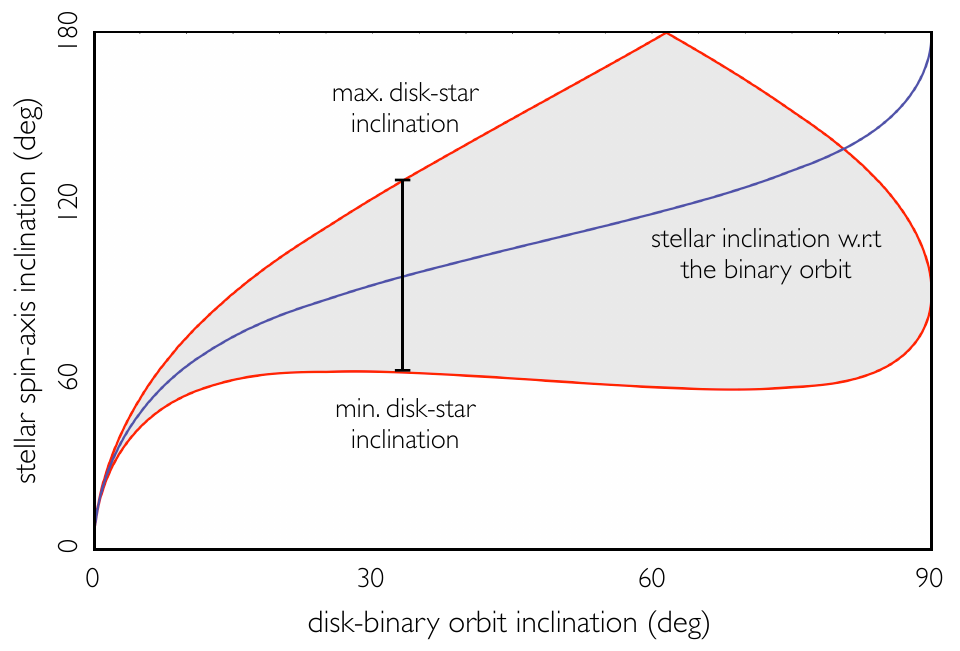}
\caption{Resonant excitation of spin-orbit misalignment. Post-resonant encounter stellar inclination of the star (measured in a frame coplanar with the binary orbit) as a function of disk-binary inclination is shown as a purple curve. Corresponding maximal and minimal spin-orbit misalignments between the stellar and the disk's angular momentum vectors, attained over a precession cycles are depicted as red curves. Note that the entire possible range of spin-orbit misalignments is attainable with a disk-binary inclination $\beta' \leqslant 65 \deg.$} 
\label{adiabat}
\end{figure}

Ultimately as the disk dissipates, $\tilde{\delta}$ approaches zero and the Hamiltonian (\ref{HammyK}) becomes a trivial one, whose flow is represented by concentric circles on the phase plane. Accordingly, the post-encounter inclination can be calculated from the definition of the adiabatic invariant:
\begin{eqnarray}
\mathcal{J} = 2\pi (1-\cos(\tilde{\beta})) = \left[ \oint \Phi_{\rm{separatrix}} d\phi \right]_{\tilde{\delta}_{\rm{crit}}},
\label{adiabatic}
\end{eqnarray}
where the separatrix equation at critical $\tilde{\delta}$ can be obtained by substituting the value of $\mathcal{K}$ corresponding to the unstable equilibrium point into equation (\ref{HammyK}). A noteworthy property of the solution (\ref{adiabatic}) is that it depends exclusively on $Z'$. In other words, \textit{in the adiabatic regime, the excitation of spin-orbit misalignment is independent of all system parameters except the primordial disk-binary plane inclination.}

Using the approach delineated above, we have mapped out the relationship between the primordial disk-binary inclination and the final (post-encounter) stellar inclination (also with respect to the binary orbital plane). This function is shown as a purple curve on Figure (\ref{adiabat}). In the decoupled $\tilde{\delta} = 0$ regime, both the stellar and the disk's inclinations, measured with respect to the binary orbit, are preserved. However, because of differential precession, the mutual disk-star inclination undergoes cyclical variations \citep{batygin2012}. The maximal and minimal mutual inclinations are obtained when the stellar orbit crosses the $y=0$ line with a negative and a positive values of $x$ respectively. The associated range of mutual inclinations is depicted in Figure (\ref{adiabat}) with red lines. As shown in the Figure, a broad array of spin-orbit angles, ranging from perfectly disk-aligned states to perfectly disk-anti-aligned star states can be produced as a consequence of passage through the secular spin-orbit resonance. 

\subsection{Magnetic and Gravitational Excitation}

\begin{figure}
\includegraphics[width=1\columnwidth]{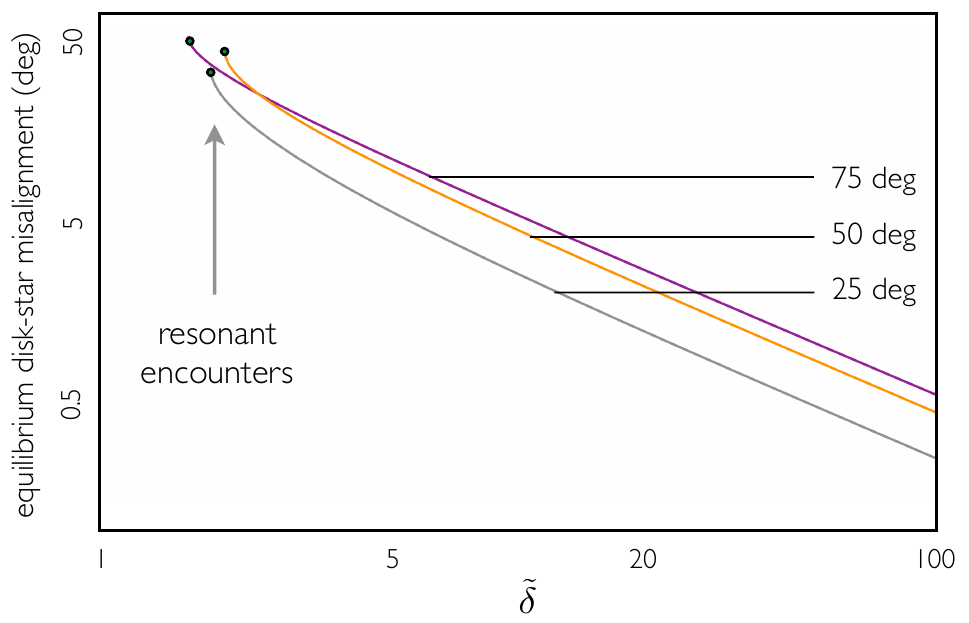}
\caption{Gravitationally enforced misalignment between an equilibrium point (i.e. initial condition) of the Hamiltonian (\ref{HammyK}) and the disk-aligned state as a function of the resonance proximity parameter $\tilde{\delta}$ for three disk-binary inclinations; 25, 50 and 75 degrees. The fact that the gravitational equilibrium does not lie directly on a disk-aligned state provides a seed inclination for magnetic torques to operate.} 
\label{forcedinc}
\end{figure}

As pointed out by \citet{lai2011} (see also \citealt{1999ApJ...524.1030L}) and rehashed in section 2.2 of this paper, a finite disk-star misalignment can be amplified by magnetic disk-star interactions. Within the context of the picture outlined above, it is tempting to assume that magnetic torques will be of no consequence prior to the encounter with the separatrix, since adiabatic invariance ensures alignment between the disk and the star. However, a more thorough examination shows that the equilibrium on which the star is envisioned to reside at $\tilde{\delta} > \tilde{\delta}_{\rm{crit}}$ deviates away from the exact disk-aligned state by a small amount\footnote{This misalignment refers to the forced component of the inclination vector (see Ch.7 of \citealt{md2000})}. Consequently, the seed misalignment, needed for the magnetic tilting process to become active is ensured to exist from purely gravitational considerations. The amplitude of this equilibrium misalignment is shown as a function of $\tilde{\delta}$ in Figure (\ref{forcedinc}) for the three choices of inclination considered above. Note that even for high values of $\tilde{\delta}$, the misalignment can be consequential (e.g. $\sim 0.5 - 5 \deg$).

\begin{figure}
\includegraphics[width=1\columnwidth]{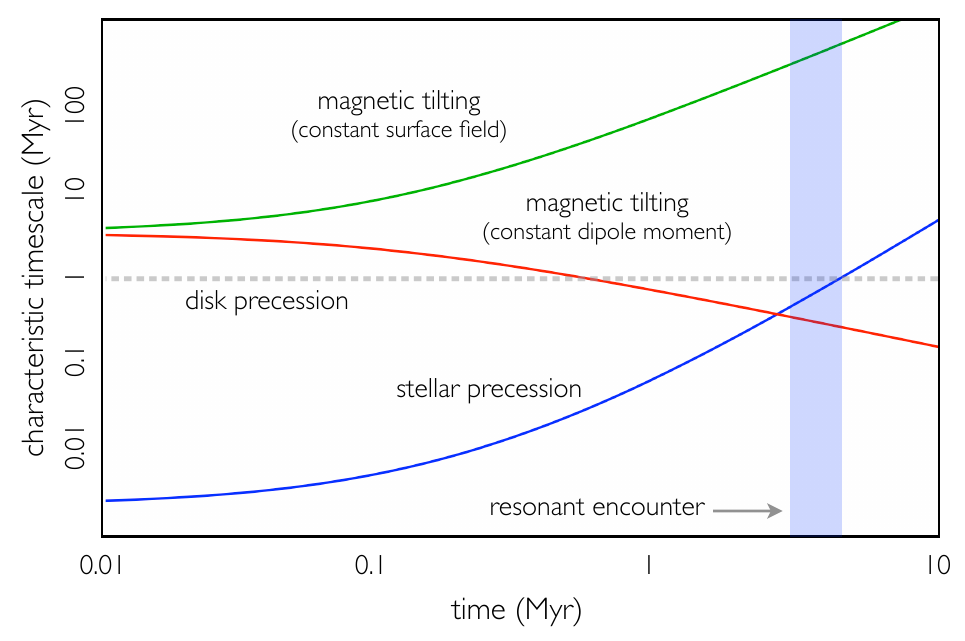}
\caption{Characteristic timescales as functions of disk age. Taking into account the physical evolution of the star and the disk, the stellar precession timescale is shown as a blue line, while magnetic tilting timescales assuming a constant surface field and a constant dipole moment are shown as green and red curves respectively. The disk precession timescale is nominally chosen to be $1$ Myr, as in the numerical simulations discussed in the text. The time interval at which the resonant encounter will take place (depending on $\beta'$) is highlighted in blue.} 
\label{timescales}
\end{figure}

The extent to which the purely gravitational picture outlined above will be altered by the incorporation of magnetic fields depends on the assumed parameters inherent to the system. This can be gathered immediately by considering the characteristic magnetic tilting timescale:
\begin{eqnarray}
\tau_{B} = \left[\frac{\zeta}{3} \frac{B_{\star}^2}{\mu_0} \frac{R_{\star}^4}{I M_{\star} \Omega_{\star} {a'}_{\rm{in}}^3}  \right]^{-1}.
\label{tauB}
\end{eqnarray}
Evaluated using the physical parameters quoted in section 2, a stellar rotation period of $\sim 5$ days \citep{affer, bouvier2013}, and a surface field of $B_{\star}\simeq 1.5$ kGauss \citep{johns2007,gregory2012}, this timescale (either assuming a constant surface field or a constant magnetic dipole moment in time), along with the characteristic stellar precession timescale, and a typical disk precession timescale are plotted over a maximal disk lifetime in Figure (\ref{timescales}).

Owing to gravitational contraction, if an evolutionary track with a constant surface field is assumed, the cumulative effect of the magnetic torque is smaller than that if a constant dipole moment is assumed. The former situation  is easily tractable within the context of the picture outlined above. First, let us imagine that system parameters are such that the secular resonance is encountered later than $\sim 0.5$ Myr after the birth of the star (i.e. after the characteristic magnetic tilting timescale becomes considerably longer than the disk lifetime). Under this assumption, the magnetic and gravitational acquisitions of disk-star misalignment occur on separate timescales and can be treated sequentially. 

It is worth recalling that the neighborhood of the (nearly) disk-aligned equilibrium of $\mathcal{K}$ (shown on Figure \ref{pspace}) is foliated in elliptical circulatory trajectories. This means that the system can acquire a non-zero value of $\mathcal{J}$ well before resonance crossing. Consequently, as $\tilde{\delta}$ approaches $\tilde{\delta}_{\rm{crit}}$, the trajectory will encounter the separatrix at at a moment when the phase-space area occupied by the inner branch of the critical curve matches that of the orbit. In other words, the resonant encounter will take place at $\tilde{\delta} > \tilde{\delta}_{\rm{crit}}$. As a consequence, the resonant excitation of spin-orbit misalignment will occur earlier in the disk's evolution and the acquired misalignment will be somewhat different. However, blunt evaluation shows that barring unreasonable estimates, the quantitative correction is essentially negligible and is of little interest (especially given the substantial uncertainties in other, more essential parameters such as $\nu$). 

Considerably more interesting dynamical behavior can be observed if a constant magnetic moment is assumed. As shown in Figure (\ref{timescales}), in this case the magnetic tilting timescale is comparable to the disk torquing time. Thus, rather than reasoning through the evolution within the framework of adiabatic theory, we resort to direct numerical integration of equations of motion, accounting for both, magnetic torques (equations \ref{pealeequns}) and gravitational torques arising from Hamiltonian (\ref{HammyK}). 

We initialize the system at the gravitational equilibrium point discussed above. The orbit is evolved for $10$ Myr, adopting the same choices of disk-binary inclination as before, in addition to an almost orthogonal configuration with $\beta' = 85 \deg$. Finally, for a more candid comparison, we ignore the dependence of $\nu$ on $\beta'$ and assume a disk precession period of $2\pi/\nu = 1$ Myr for all simulations. The physical evolutions of the star and the disk are assumed to proceed as described in section 2.1.

\begin{figure}
\includegraphics[width=1\columnwidth]{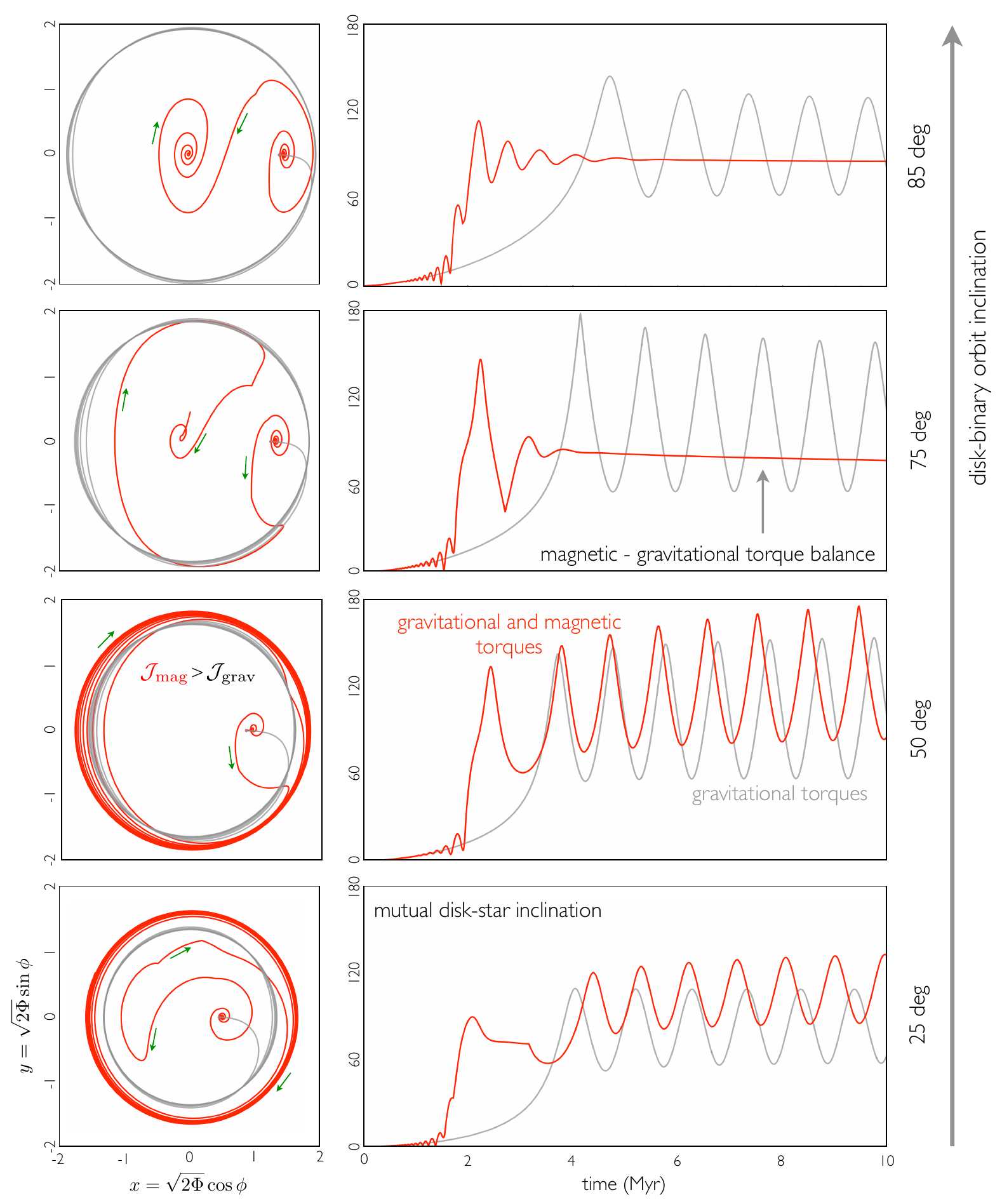}
\caption{The results of numerical integration of equations of motion. The right panels show mutual disk-star inclination as functions of time, while the left panels show the phase-space trajectories of the stellar spin axis. The red curves denote solutions that account for gravitational and magnetic torques, assuming a constant dipole moment. Meanwhile, the gray curves show solutions where only gravitational torques have been retained. While the latter adhere to the analytic solutions obtained within the framework of adiabatic theory, the former show qualitative deviations from purely conservative behavior.} 
\label{evolution}
\end{figure}

The results of the integrations are shown in Figure (\ref{evolution}), where the full integrations are plotted in red and solutions that only account for gravitational torques are plotted in gray for comparison. Immediately, a number of interesting features can be observed. First, in all simulations, the magnetically facilitated growth of $\mathcal{J}$ is evident, and the impulsive excitation of spin-orbit misalignment occurs substantially earlier when magnetic tilting is taken into account. 

For lower inclinations (i.e. $\beta' = 25 \deg$ and $\beta' = 50 \deg$), it is clear that magnetic tilting complicates the trajectory, although the qualitative behavior of the dynamical evolution is similar to that of the purely gravitational calculation. That is to say that the orbit evolves towards a quasi-circular state (in phase-space) as the disk dissipates. However, unlike the purely gravitational case, here the value of $\mathcal{J}$ continues to grow, even after separatrix crossing. This is largely due to magnetic torques and arises from the fact that a fraction of the orbit resides at $\beta < 90 \deg$. The plots of stellar inclination relative to the disk give a physical picture of the above process. 

When star-disk inclinations reach values of $\beta_i \geq 90 \deg$, the magnetic contribution in our model becomes null and the only effect is that of secular gravitational interactions. This remains true until the oscillatory trajectory brings the star-disk inclination to $\beta_i \leq 90 \deg$. At this point, the magnetic influence returns and repels the stellar inclination back to values of $\beta_i \geq 90 \deg$. This causes the inclination to oscillate in a similar fashion to the purely gravitational case, but with its trajectory is eventually excluded from $\beta_i \leq 90 \deg$. It is worth noting that a quantitative description of this effect would be significantly altered if there is in fact some magnetic influence (not considered here) for $\beta_i \geq 90 \deg$. 

The evolution at higher inclinations is qualitatively different, as the orbit evolves towards a fixed point (characterized by a balance between gravitational and magnetic torques) after crossing the separatrix. The $\beta' = 85 \deg$ case is particularly striking, as the phase-space plot depicts an initial condition that behaves as a repeller of the trajectory and the binary aligned equilibrium point serves as an attractor. It is interesting to note that similar behavior can be obtained by augmenting the Hamiltonian evolution with dissipative effects of accretion (see \citealt{batad} for an in-depth discussion).

These results imply that provided an advantageous prescription for the ingredients inherent to the magnetic torquing part of the calculation, the dynamical evolution of the system can be qualitatively altered. However, it is important to note that magnetic effects do not obstruct the acquisition of spin-orbit misalignment within the framework of the disk-torquing model but instead act to accelerate it. In turn, gravitational effects provide the root inclination needed for the magnetic torques to operate. Consequently, it seems reasonable to conclude that the magnetic torquing mechanism proposed by \citet{lai2011} may play an important, but nevertheless secondary role in explaining observed hot Jupiter spin-orbit misalignments. 

\section{Discussion}

In this work, we have considered the excitation of spin-orbit misalignment within the context of disk-torquing theory \citep{batygin2012}, taking into account magnetically-facilitated tilting of the star \citep{lai2011}. While this study builds on the previous work of \citet{batad}, it differs in two important ways. First, the treatment of gravitational torques employed in this work does not assume small inclinations and allows us to self-consistently explore the process of secular spin-orbit resonant encounters. Second, this work includes additional physics stemming from magnetic disk-star interactions \citep{1999ApJ...524.1030L}. Cumulatively, the results of our work can be summarized as follows.

Taking advantage of the separation of dynamical and physical evolution timescales inherent to the problem, we utilized adiabatic theory \citep{1991pscn.proc..193H} to analytically compute the impulsive excitation of spin-orbit misalignment during resonant encounters. The attained results suggest that \textit{the entire possible range of spin-orbit misalignments can be produced exclusively by the disk-torquing mechanism given a disk-binary orbit inclination $\beta' \leqslant 65 \deg$.} Moreover, as long as the resonant encounter takes place in an adiabatic regime, the attained inclination depends only on $\beta'$.

The inclusion of magnetic effects complicates the purely gravitational picture on a quantitative level. Primarily, magnetic perturbations drive the system through secular spin-orbit resonance at an earlier epoch, thereby leading to somewhat enhanced disk-star inclinations. At high disk-binary orbit inclinations, magnetic torques may also drive the system towards an equilibrium that corresponds to a near-alignment between the stellar spin-axis and and the binary orbit angular momentum vector. However, we note that in order to obtain significant deviations away from a purely gravitational solution, we made favorable (and perhaps unrealistic) assumptions about the strength of the field \citep{donati2010,gregory2010} and the disk truncation radius \citep{mattpud}. Consequently, it may be true that the aforementioned corrections are not too relevant in reality. Either way, the capacity of the disk-torquing model to explain the origins of spin-orbit misalignments of hot Jupiters, within the context of smooth disk-driven transport is not hindered by the inclusion of additional physics. 

An obligatory property of the disk-torquing model considered in this work is the prevalence of stellar companions during the early epochs of planet formation \citep{batygin2012}. This constraint is not as stringent as that inherent to (for example) the Kozai migration model \citep{wumurray,fabtremaine,naoz} which requires longer-lived binaries than the model presented here. As such, we would expect the disk-torquing scenario to play out more frequently across a given sample of planetary systems than Kozai migration simply owing to the greater frequency of short-lived relative to long-lived binary systems.  Although a significantly elevated fraction of multi-stellar systems in young star clusters is observationally established \citep{ghez1993,kraus2011}, it seems natural to additionally expect a corresponding correlation between hot Jupiter spin-orbit misalignments and the existence of present-day wide-orbit companions.

To this end, the observational survey of \citet{2013arXiv1312.2954K} has not found a statistically significant parallel between the two measurements. However, in interpreting these results, it is important to keep in mind that the dynamics of stellar clusters can be extremely complex (see e.g. \citealt{2007MNRAS.378.1207M,adams2010}), and dissolution of multi-stellar systems as well as binary exchange reactions will act to obscure a direct relationship between primordial and present (i.e. cluster-processed) field stellar multiplicity. However, such interactions may well be specific to high density clusters (\citealt{Duchnk}) and so additional computational effort is required to determine whether the theory presented here is consistent with observations of stellar multiplicity. This issue should be examined in detail as an integral component of a future study.

An observational trend that our model does not explicitly account for is the dependence of hot Jupiter misalignments on the effective temperature of their host stars \citep{winn2010}. Although, an explanation that invokes the mass-dependence of tidal dissipation for why predominantly hotter stars ($T_{\rm{eff}}\gtrsim 6250$K) are characterized by large obliquities has been presented \citep{winn2011,lai2012,albrecht2012}. Within the context of the envisioned scenario, all hot Jupiters originate with high orbital obliquities, and spin-orbit misalignments are subsequently erased by tidal dissipation preferentially in low-mass stars. 

Generally, the disk torquing model discussed in this work does not preclude subsequent, additional effects owing to tidal dissipation. However, in light of the recent criticism of this narrative by \citet{rogerslin2013}, it may be worthwhile to speculate about an alternative scenario. As already mentioned above, if disk-driven migration is the dominant mode of early orbital transport, the generation of spin-orbit misalignments requires a wide-spread existence of binaries in birth clusters. It has been noted observationally that stellar binarity (and the stellar orbital distribution) are both strong functions of stellar mass \citep{kraus2011} with the trend being to increase binarity with higher $T_{\rm{eff}}$ systems, mirroring the observed $T_{\rm{eff}}$-misalignment trend. Consequently, a handle on the observed misalignment-$T_{\rm{eff}}$ correlation may conceivably be obtained by further examining the tally and the longevity of multi-stellar systems in star-formation environments as a function of their mass. While a potentially fruitful avenue of reasoning, additional observational and modeling efforts will be required to definitively evaluate this possibility. \\

\textbf{Acknowledgements} We thank the anonymous referee for a careful review of the paper which led to an enhanced manuscript. During the review of this paper, we have become aware that Lai (2014) arrived at similar results simultaneously and independently. K. Batygin acknowledges the generous support from the ITC Prize Postdoctoral Fellowship at the Institute for Theory and Computation, Harvard-Smithsonian Center for Astrophysics. C. Spalding acknowledges the generous support from the CONOCO Graduate Fellowship in Geology at the California Institute of Technology.

\end{document}